\documentclass{llncs}

\usepackage{html}
\usepackage{url}
\usepackage[abbr]{harvard}
\usepackage{amssymb}
\usepackage{graphicx}
\usepackage[cmex10]{amsmath}
\usepackage[caption=false,font=footnotesize]{subfig}
\usepackage{acronym}
\usepackage{epstopdf}

\acrodef{PTDF}[PTDF]{Power Transfer Distribution Factor}
\acrodef{CNA}[CNA]{Complex Network Analysis}
\acrodef{UCTE}[UCTE]{Union for the Coordination of Transport of Electricity}
\acrodef{BA}[BA]{Barab\'{a}si-Albert}
\acrodef{ER}[ER]{Erd\H{o}s-R\'{e}yni}
\acrodef{LCC}[LCC]{Largest Connected Component}
\acrodef{CLM}[CLM]{Crucitti-Latora-Marchiori}
\acrodef{DC}[DC]{Direct Current}

\newenvironment{keywords}{
       \list{}{\advance\topsep by0.35cm\relax\small
       \leftmargin=1cm
       \labelwidth=0.35cm
       \listparindent=0.35cm
       \itemindent\listparindent
       \rightmargin\leftmargin}\item[\hskip\labelsep
                                     \bfseries Keywords:]}
     {\endlist}

\begin{document}

\mainmatter

\title{Context-Independent Centrality Measures\\Underestimate the Vulnerability of Power Grids}

\titlerunning{Vulnerability of Power Grids}

%
%
\author{Trivik Verma\inst{1,2,3} \and Wendy Ellens\inst{2} \and Robert E. Kooij\inst{1,2}}

\institute{Delft University of Technology, EEMCS, The Netherlands
\and
Performance of Networks and Systems, TNO, The Netherlands
\and
Institut f. Baustoffe (IfB)/ETH Risk Center, Zurich, Switzerland } 

\maketitle

\begin{abstract}
Power grids vulnerability is a key issue in society. A component failure may trigger cascades of failures across the grid and lead to a large blackout. Complex network approaches have shown a direction to study some of the problems faced by power grids. Within Complex Network Analysis structural vulnerabilities of power grids have been studied mostly using purely topological approaches, which assumes that flow of power is dictated by shortest paths. However, this fails to capture the real flow characteristics of power grids. We have proposed a flow redistribution mechanism that closely mimics the flow in power grids using the \ac{PTDF}. With this mechanism we enhance existing cascading failure models to study the vulnerability of power grids.

We apply the model to the European high-voltage grid to carry out a comparative study for a number of centrality measures. `Centrality' gives an indication of the criticality of network components. Our model offers a way to find those centrality measures that give the best indication of node vulnerability in the context of power grids, by considering not only the network topology but also the power flowing through the network. In addition, we use the model to determine the spare capacity that is needed to make the grid robust to targeted attacks. We also show a brief comparison of the end results with other power grid systems to generalise the result.
\end{abstract}

\begin{keywords}
power grids, complex networks, cascading failures, flow redistribution, centrality, tolerance parameter, vulnerability
\end{keywords}

\section{Introduction}
Power grids are the backbone of modern society. A power grid is a network of generation, transmission and distribution sub-stations that transmit power to households and local businesses. Energy is transferred through high-voltage transmission lines in order to reduce losses over large distances. The evolution of high-voltage grids has been influenced by social, technical, economic, political and environmental decisions that favour economic prosperity, security and quality of life. A disruption in the functioning of power grids can have severe impact on the social welfare of society \cite{Bompard3}. The reliability of power grids is thus very critical for optimal functioning of society. 

In recent decades a lot of blackouts have been experienced by power grids across the world \cite{McLinn}. Power outages are consequences of perturbations that overload the entire system by spreading flows across the network. These perturbations range from severe weather conditions to several human errors. Failure of a component in the power grid may cause some power to be redirected to its neighbours, which in turn may get overloaded and redirect power to \emph{their} neighbours causing them to fail. These failures may initially start with any component of the power grid (faults at power station, transmission lines, defects in the distribution network or even a small short-circuit) that propagates its effects to its direct neighbours and so on. This effect is called a \emph{cascading failure}. 

Like most complex networks, power grids are also made up of small components that act in a simple manner and together give rise to a behaviour that emerges complex in nature \cite{Mitchell}. As the size of power grids is increasing over the years, it is becoming very important to understand the emergent behaviour of such systems. The question still remains how complex and vulnerable these systems are and how the small simple components give rise to a larger complex mesh of power. Analysing vulnerability of power grids is a significant issue in society and much has been contributed towards such efforts \cite{Albert1,Bompard,Bompard1,Bompard2,Bompard3,Bompard4,Crucitti,Kinney,Koc,Simonsen}. These discussions give rise to an interesting direction of research that not only deals with topological (purely relating to the structure and interconnection pattern of the grid) vulnerability but also looks into the load flows of power grids \cite{Arianos,Crucitti,Kinney,Nasiruzzaman,Youssef}.

Disruptions (accidents, human errors, component failures, etc.) in a power grid may lead to a failure that propagates through the grid and results in a large-scale power outage. This cascading failure phenomenon has been researched extensively \cite{Bompard,Crucitti,Kinney,Pahwa,Simonsen,VanEeten,Youssef} but many properties of power grids are still missing from the analyses and need to be included for a more accurate study. This leads us to study cascading failures by incorporating properties of flows in power grids in order to deal with the problem of power outages in high-voltage power grids. Through this study we want to identify the most \emph{vulnerable} nodes in high-voltage power grids (whose removal causes most damage to the grid) and find out how much capacity is required at each node so that the power grid sustains attacks and failures. More concretely we want to answer two questions: 1) which measure of centrality in complex networks gives the best indication of node vulnerability in power grids and 2) what should be the minimal node \emph{tolerance level} (node capacity relative to the typical loading level) in order to prevent cascading failures that have impact on more than ten percent of the energy delivery.

\begin{figure}
\begin{center}
\includegraphics[width = .8\textwidth]{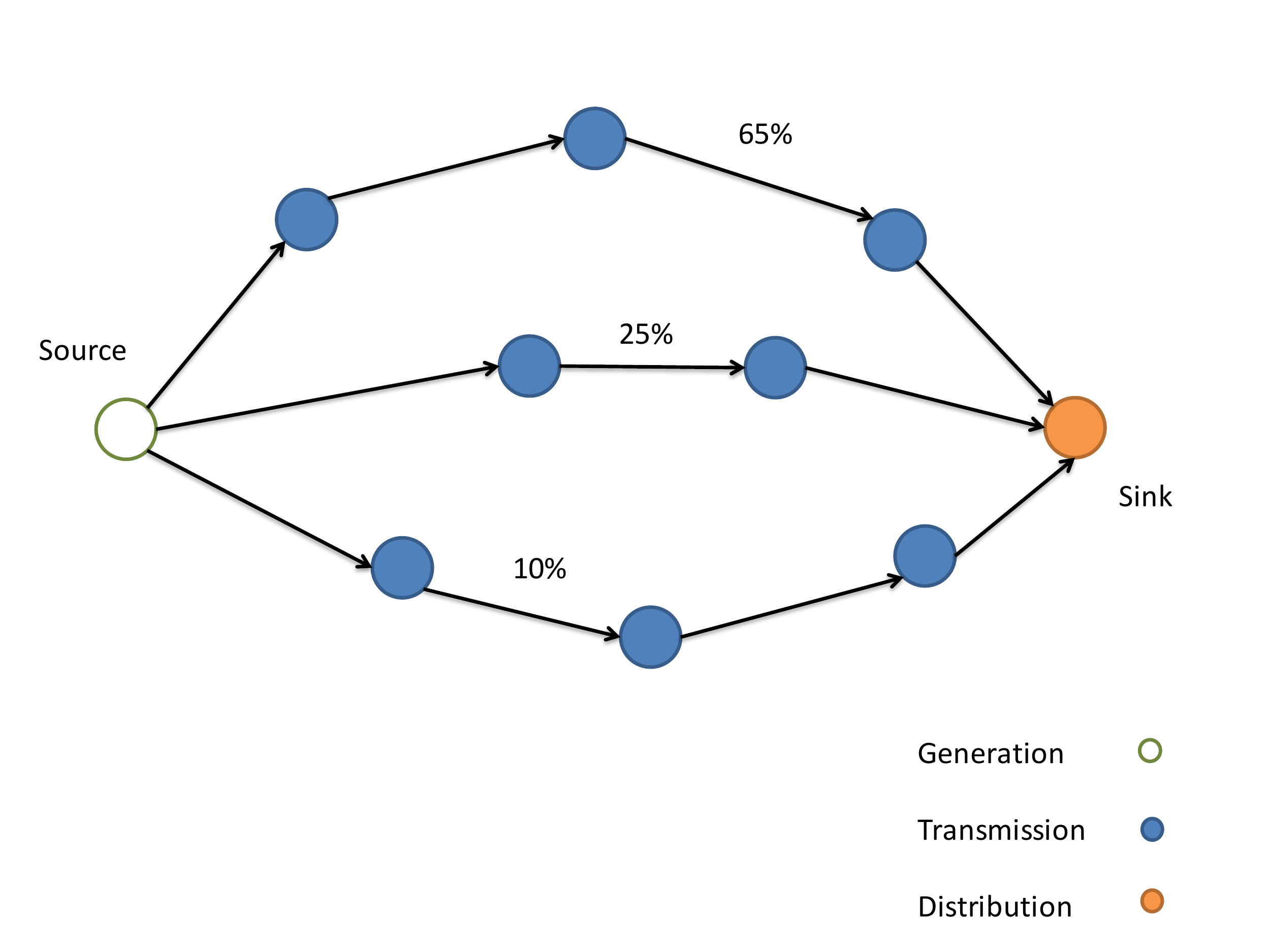}
\caption{A network diagram showing the distribution of power across multiple paths between a pair of generation and distribution nodes of a power grid.}
\label{fig:NetworkFlowDiagram}
\end{center}
\end{figure}

To answer these questions, we need to predict the evolution of cascading failures in a high-voltage power grid with more accuracy than existing theoretical \ac{CNA} models. The existing cascading failure models \cite{Crucitti,Kinney} assume that power flows between two nodes via the shortest path that exists between them and a few others deal with \ac{DC} flow analysis incorporating a dynamic behaviour in the model \cite{Koc,Youssef}. It is of little significance whether the length of the shortest path is based on distance or impedance of each link in a path. Power, in reality, flows through multiple paths that exist between two nodes and is divided between these paths based on electrical properties (impedance, conductance, capacitance) of the transmission links as depicted in Fig. \ref{fig:NetworkFlowDiagram}. Hence, we propose a model with a more realistic flow redistribution mechanism and use it to simulate cascading failures. We trigger a cascading failure by removing a node with the highest centrality \cite{Boccaletti} according to four different centrality measures to find out which removal is the most disruptive and to know which measure best captures the vulnerability of nodes to cascading failures. Thereafter, in the second phase, we perform a new set of cascading failure simulations for different values of the tolerance level by deleting the most vulnerable node according to the centrality in the first phase. This second set of simulations is aimed at finding the minimal tolerance level needed to prevent high-impact cascading failures.

The next section, Sect. \ref{sec:Literature}, talks about the context in which this work is carried out. In Subsect. \ref{sec:PastWork} we describe literature in which power grids are studied from a complex networks point of view, while Subsect. \ref{sec:PTDF} explains the basis for our proposed flow redistribution mechanism. In Sect. \ref{sec:Model}, we describe our cascading failure model and its constituent parts in detail. Section \ref{sec:Results} is about the simulations. We start by defining the settings for our simulations. Next, we talk about the performance measures that are used to quantify the vulnerability of power grids in case of cascading failures. The section continues with a description of the simulation approach and finishes with a discussion of the simulation results. Section \ref{sec:Conclusions} marks the end of this article with a short summary and focal points of this research. It also explains future potential that emerges from this work. 

\section{Related Work}\label{sec:Literature}

\subsection{Complex Network Analysis}\label{sec:PastWork}

This section is dedicated to the relevant work carried out in the field of power grids and cascading failures. We describe summaries of some relevant papers in detail and focal points of each work that might be useful for expanding our model.

Researchers initially focussed on static failures in complex networks \cite{Albert1,Albert2,Callaway,Cohen}. These include the removal of certain components of a network and thereby evaluating the drop in performance of such a network. However, many infrastructure networks like the transport grid/gas pipelines may lose a sizeable part of their performance after a static failure and cause an overload in other parts of the network leading to a partial or complete disruption of facilities. This redistribution of flows led to development of dynamical approaches \cite{Crucitti,Motter} and their application to more real-world networks.  

Next we talk about a dynamic model of cascading failures for complex networks: \ac{CLM} model, by \citeasnoun{Crucitti}. They build a simple model to show how small triggers could lead to catastrophic events in infrastructure networks. The model explains redistribution of flows within the network upon break down of a single node. They show that the collapse of a single node with the highest load is sufficient to considerably lower the efficiency of the entire network. 

The \ac{CLM} model has been adapted to the North American high-voltage power grid by \citeasnoun{Kinney}. The north American power grid is one of the most complex technological networks and transmits electricity but also failures to many parts of the grid. They modelled the grid with certain assumptions about how the load is transferred from one transmission substation to another. Their results suggest that a breakdown of transmission sub-stations could lead up to a loss of $25\%$ of efficiency by spreading flows across the network. Furthermore, they found out that disruption of $40\%$ of the transmission sub-stations lead to cascading failures. 

Considerable work has been carried out by researchers in the field of cascading failures in the direction of stationary network load models. The above mentioned studies have that feature in common. \citeasnoun{Simonsen} have taken a step further and studied the transient dynamics of power systems caused by an initial shock in the network. They have justified their work by the fact that loads have oscillations while adjusting to the new structure of the network and some parts may breakdown before the network stabilises again. The lack of a dynamic load model greatly overestimates the robustness of the network. Furthermore, they conclude that using stationary network load models might predict a different failure sequence for the network far from reality. 

Until now we have discussed literature that marks direct application of context-independent concepts and measures from \ac{CNA} to power grids. Power grids cannot be encapsulated by generic properties of a complex network. \citeasnoun{Bompard4} have done major work in exploring the shortcomings of generic application and designing new methodologies to study power grids more efficiently. 

In \cite{Bompard} they analyse the structural vulnerabilities of power grids by extending the topological approach to include the physical operative state of the grid in terms of the flows distributing in the links. The authors designed new metrics such as \emph{entropic degree} and \emph{net-ability} and compared them with the traditional purely topological metrics. Entropic degree takes into account not only the number of connections a node has but also the distribution of weights around these connections. It provides a fairly quantitative measurement of the importance of sub-stations in a power grid. Net-ability describes how efficiently power flows over the lines of a transmission grid from supply to demand side. They further extended the topological approach to incorporate several other features of a power grid such as line flow limits, flow paths and demand/supply distribution \cite{Bompard1}. \emph{Path redundancy} quantifies the available redundancy of paths in transmitting power from generation to distribution sub-stations using the concept of entropy. In case of a cascading failure, \emph{Survivability}, based on path redundancy and net-ability, measures the effectiveness of a grid to match demand with the generation. 

On the lines of extending topological approach to analyse power grids, \citeasnoun{Bompard2} formulated an electrical betweenness based approach that captures the specific properties of power grids in more detail. This metric focusses on the impedances of the transmission lines. They compared this with the traditional betweenness metric for IEEE bus networks and found that this traditional centrality measure underestimates the vulnerability of power grids.

Centrality is the backbone for identifying critical components in a complex network. \citeasnoun{Tutzauer} proposed a centrality measure based on entropy that characterises networks with flow along multiple paths. His motivation lies in the idea that centrality measures must be related to the underlying network properties. Very often, researchers use centrality measures to validate models for a particular type of network that may have flow characteristics different from the ones assumed by the measure. \citeasnoun{Borgatti} points out that this implicit assumption about the flow characteristics of network may give rise to misleading results. 

\citeasnoun{Koc} have designed a context-dependent centrality measure to assess the importance of nodes in a power grid. The measure describes the significance of a node based on the amount of power flowing through that node to its neighbours. \emph{Node Significance} is a directed and weighted measure and a good estimate of the robustness of a node in a power grid. 

Bompard's seminal work may have inspired optimisation of distributed flows in power grids from a design perspective. \citeasnoun{Asztalos} used an edge weighting scheme to optimise the flow efficiency and robustness of scale-free networks to cascading failures. They studied models where flow is distributed and initiates from all paths between a pair of source and sink nodes.

\citeasnoun{VanEeten} illustrate the importance of energy networks. They carry out empirical investigations in the interdependencies of critical infrastructures. Their work involves analysing data from news reports and correlations between cascading failures across different infrastructures. Their findings were rather focused on two domains: telecommunications and energy. Most cascades have propagated from energy networks to telecommunications and there is no sign of a bidirectional relationship. 

Power grids have been around for almost two centuries and the continuing complex interconnection patterns have made them more liable to disruptions. \citeasnoun{Pahwa} have explored mitigation strategies against cascading effects and agreed upon the use of distributed renewable energy sources for decreasing the load on power grids. Subsequently, \citeasnoun{Youssef} have also explored robustness measures of power grids with respect to cascading failures using a \ac{DC} power flow model. 

\subsection{Power Flows}\label{sec:PTDF}
In a power grid, flow between a pair of source and sink nodes does not follow the shortest path between them. Power follows a distributed path based on many system properties (impedance, conductance, capacitance) that impose certain limits on the capacity of these paths \cite{Bompard}. Every path that constitutes the flow from source A to sink B is made of several links each. These links have different parameters and each path may contribute a different proportion of power to the flow from source A to sink B. 

The amount of information flowing between a pair of nodes is reflected by the shortest path between the nodes \cite{Albert1,Crucitti,Rosas-casals}. Contrary to this assumption, in power engineering, electric current (the information flowing in power grids) follows multiple paths between a pair of nodes. Merely using simplified measures from the theory of \ac{CNA} for defining flow redistribution in a power grid is not enough. The flow in a power grid can be represented in essence by the \ac{PTDF}.

The \acf{PTDF} \cite{Dobson} describes the sensitivity of each transmission link to power injection at a particular node and withdrawal at a reference node. This reference node is also known as \emph{Slack}. If the slack node is missing, then a generator node with the highest real power is chosen as slack for the system. 

\ac{PTDF} of a component T for a flow between source A and sink B reflects the percentage of power that flows through that component T. For instance, when component Z (part of a path between A and B) has a PTDF of $0.2$ for a transfer from A to B, then a transfer of 100 MW from A to B would result in an increase of 20 MW at component Z. On the other hand, if the same factor is $-0.2$, then this transfer would result in a decrease of 20 MW at component Z. This implies that for calculating the \ac{PTDF} of a transmission link, the source and sink nodes must be specified. In the absence of a sink, the reference node acts as the sink that demands the injection at source in the first place. \ac{PTDF}s are very useful for analysing the change in load at a particular transmission link and identifying the links that might overshoot their capacity. 

For a network with $N$ nodes and $K$ transmission links, the \ac{PTDF} matrix is given by, 
\begin{equation}\label{eq:PTDFMatrix}
A =
\left[ {\begin{array}{cccc}
 a_{11} & a_{12} & \ldots & a_{1N} \\
 & & \vdots & \\
 a_{K1} & a_{K2} & \ldots & a_{KN} \\
\end{array} } \right] \quad ,
\end{equation}
where $A$ is a $K \times N$ matrix and $a_{ij}$ is the change in power of link $i$ when 1 unit of power is injected at node $j$ and withdrawn at the slack node. 

The \ac{PTDF} of a transmission link $i$ for a flow between a pair of nodes $g$ and $d$ can be defined using \eqref{eq:PTDFMatrix} and the slack node as follows, 
\begin{equation} \label{eq:PTDFLine}
\rho^{gd} = a_{ig} - a_{id} \quad ,
\end{equation}
where $\rho_i^{gd}$ is the change in power of link $i$ when 1 unit of power is injected at node $g$ and withdrawn at node $d$.

Computation of the \ac{PTDF} matrix requires a model to calculate load flows and in most cases a \ac{DC} model of power flow \cite{Dobson} is used. \ac{DC} power flow models are widely used by the power engineering sector to compute load flows for high-voltage transmission grids. Their reliability has been long questioned and not completely accepted for all application purposes but it is nonetheless a good approximation of active power flows and usually gives values close to $\pm 5\%$ \cite{Purchala} as long as the conditions forming the basis of the method are met with. For a high voltage transmission grid, the conditions for a \ac{DC} model are satisfied \cite{Purchala}. 

\section{Model}\label{sec:Model}
\begin{figure}
\begin{center}
\includegraphics[width = .8\textwidth]{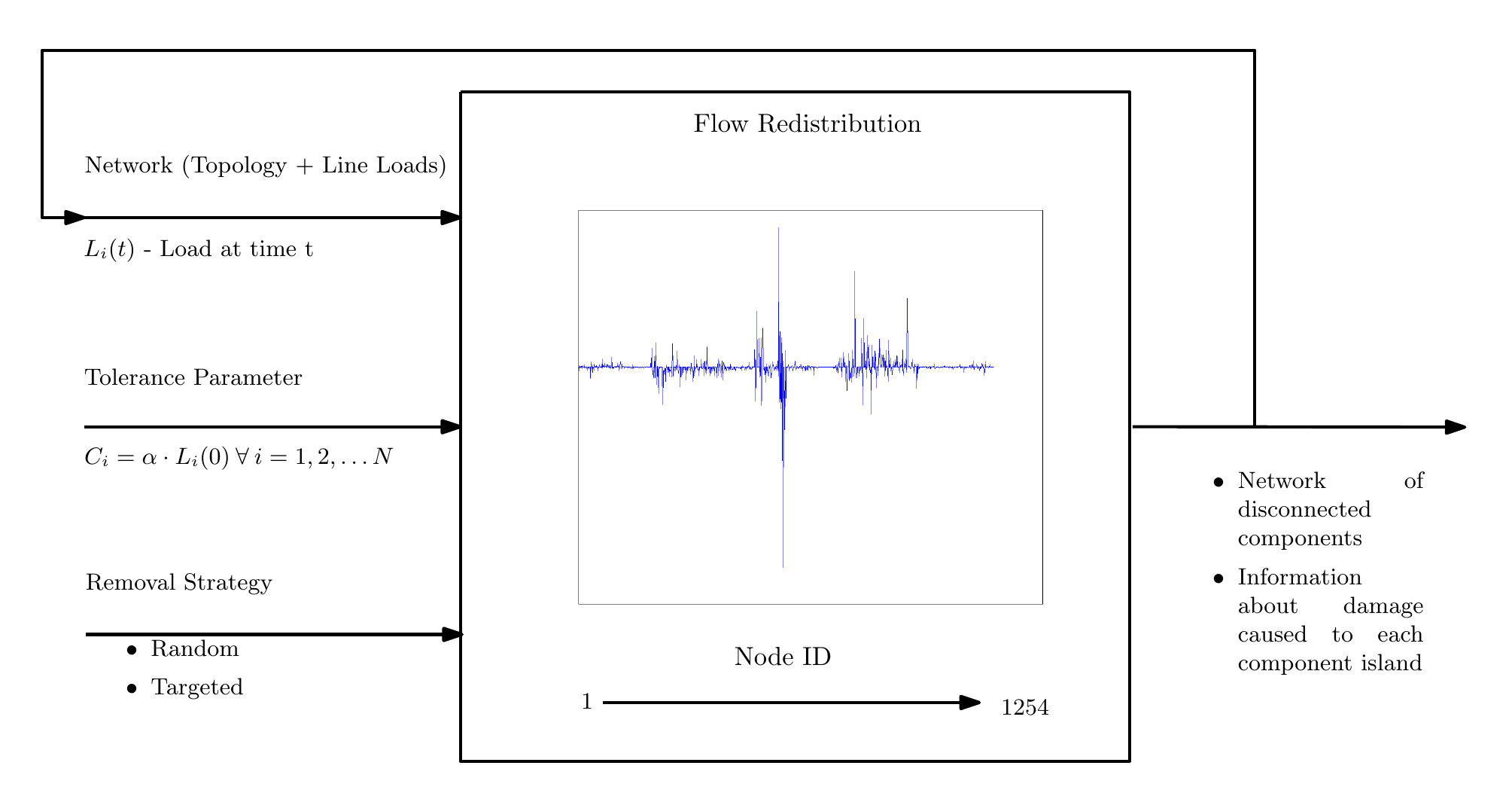}
\caption{Scheme showing the constituent parts of the cascading failure model. Left is the input, in the middle is the flow redistribution and to the right is the output. The process is iterated until no more changes, i.e. there is no more flow redistribution and no more nodes are removed because of overloading.}
\label{fig:Model}
\end{center}
\end{figure}

The cascading failure model that we have designed for simulating a high-voltage power grid has three distinct parts, namely input, redistribution and output illustrated in Fig. \ref{fig:Model}. We explain these parts in the following three subsections.

\subsection{Input}
The input settings for our model consists of the  tolerance parameter, the network model and the removal strategy.\\

\underline{Tolerance parameter $\alpha$}\\
Capacity is a characteristic property of each node/link in any network and it dictates the maximum load a node/link can withstand. Naturally, we will assume that capacity $C_i$ of a node $i$ is directly proportional to its initial stable load $L_{i}(0)$ \cite{Motter} at the start of the simulation (time $t = 0$). It is defined as follows, 
\begin{equation}\label{eq:Alpha}
C_i = \alpha \cdot L_i(0) \qquad \forall i = 1,2,\ldots,N\, \quad ,
\end{equation} 
where $L_i(0)$ is the current power flowing through each node at the moment when a \ac{DC} power flow solution was calculated. This proportionality coefficient $\alpha \geq 1$ is the tolerance parameter that marks an upper bound on the load that can flow through each node at any time during the simulation. $\alpha$ is technology dependent, i.e. it cannot be an unreasonably high value. 

From \eqref{eq:Alpha}, it is evident to state the following relation between the tolerance parameter and loading level which is a more inherently natural concept to grasp,
\begin{equation}\label{eq:LoadingLevel}
\alpha = \frac{100}{ll} \quad ,
\end{equation}
where,\\
$\alpha$ is the tolerance parameter and $ll$ is the loading level expressed as a percentage. 

The relation in \eqref{eq:LoadingLevel} implies that when the loading level is say $50 \%$, the tolerance parameter will be $\alpha = 2$, i.e. the network can handle twice its initial load at each node. \\

\underline{Network Model}\\
High-voltage transmission grid data is used \cite{UCTE}. It has $1254$ nodes and $1944$ directed links. The network has the following characteristics, \\
$average\:clustering\:coefficient\;<C> = 0.0126$ \\
$diameter\;d = 48$ \\
$average\:path\:length\;<l> = 17.28$

\begin{figure}
\begin{center}	
	\subfloat[Degree Distribution]{
	\includegraphics[width=.6\textwidth]{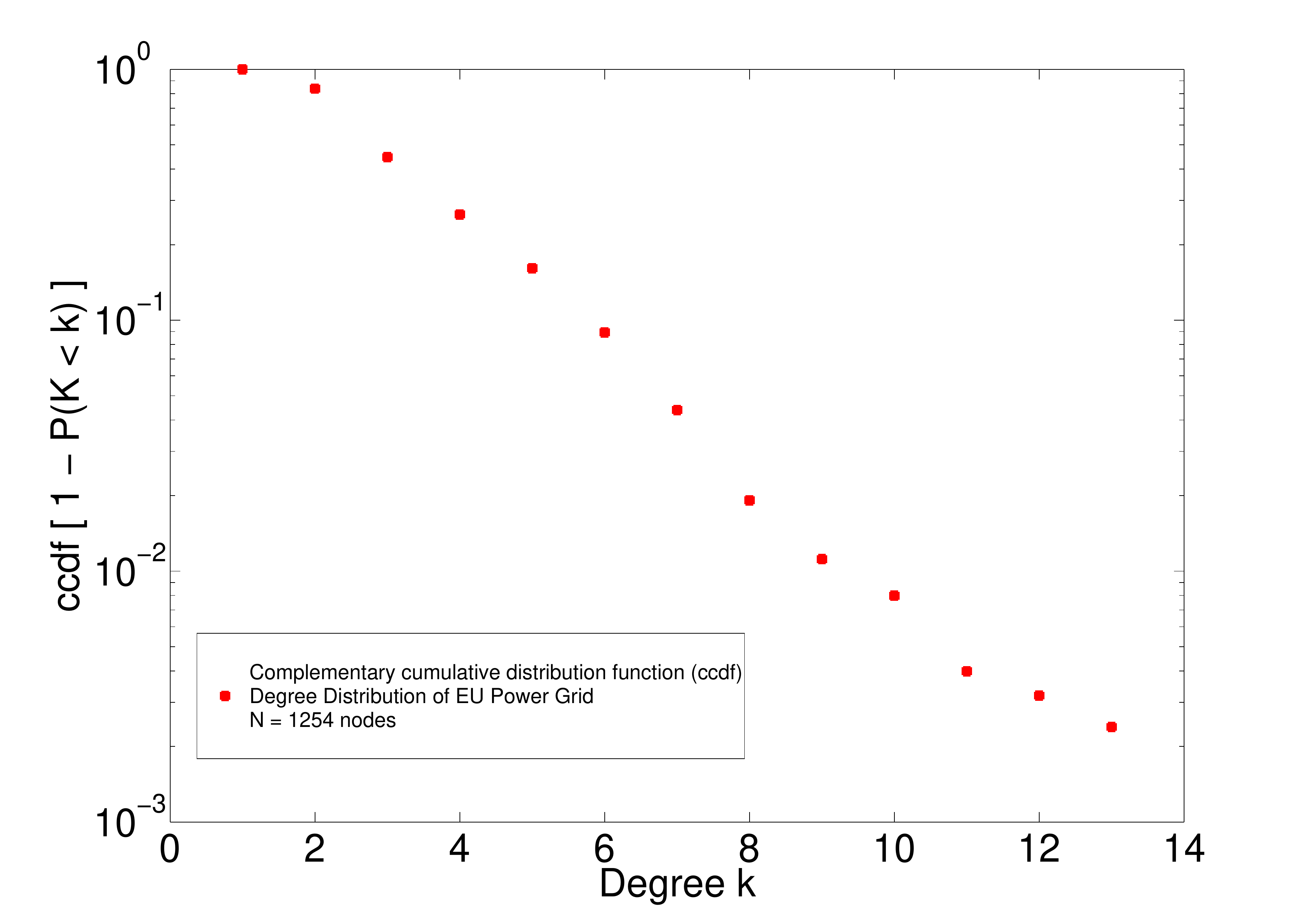}
	\label{fig:figure1}
	}
	\subfloat[Betweenness Distribution]{
	\includegraphics[width=.6\textwidth]{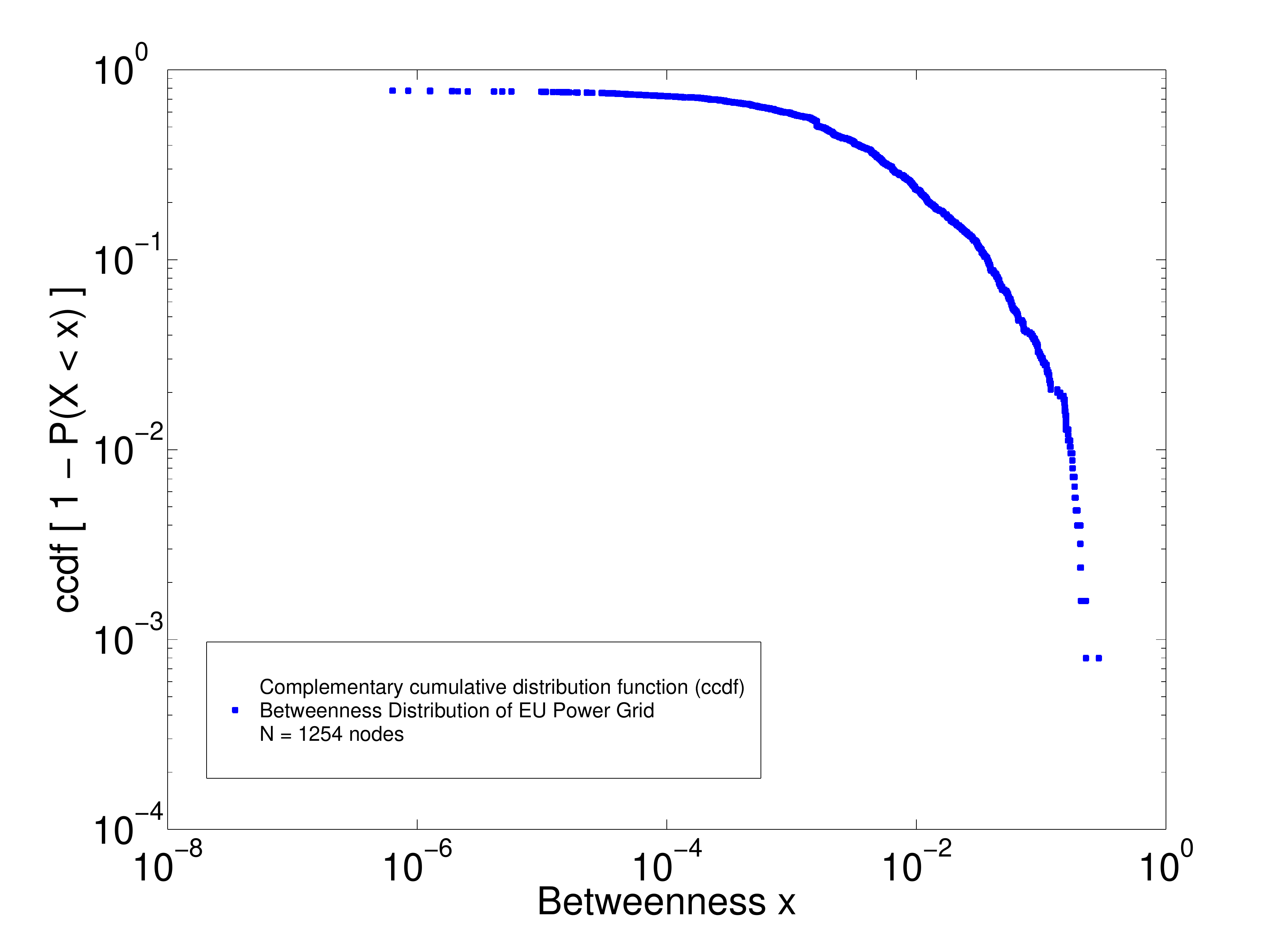}
	\label{fig:figure2}
	}
\caption{Degree and Betweenness distributions of EU power grid. The topology of this network is considered undirected and unweighted for these distributions.}
\label{fig:Distributions}
\end{center}
\end{figure}

The degree and betweenness distributions of this network are plotted in Fig. \ref{fig:Distributions}.

This network data consists of the topology of the network, the demand and supply parameters at each generation and load station, and node voltages. A \ac{DC} power flow solution with all these parameters gives an approximate amount of power flowing through all the high-voltage transmission links of the grid. These load values act as weights in our analysis. This is a stable snapshot of the system that we use for future modelling and simulation purposes.  

The \ac{DC} power flow solution comes from a MATLAB package called MATPOWER \cite{Matpower}. The calculation of \ac{PTDF} matrices is also a part of this package. It is used for modelling the power grid as a network to solve power flow equations and carry out other power flow computation problems. \\

\underline{Removal Strategy}\\
The cause for redistribution of flows in a power grid is due to a component failure or shut-down. In order to simulate the network for cascading failures we have to remove a node from the network to trigger the redistribution of flows. This removal can either be \emph{Random} or \emph{Targeted}. By removing the most \emph{important} (centrality \cite{Boccaletti}) node from the network we can estimate the worst case damage caused to a power grid due to cascading failures. Nodes of the power network are considered to shut down irreversibly, i.e. once they are switched off during the simulation of a cascading failure (whether due to a failure being propagated or a safeguarding strategy to cut-off a part of the network) they are assumed to be dead for the remaining period of the simulation. We made this choice because of the fact that this work aims at a short interval of time which is conceived from a fault in the network, resulting in a power outage in a matter of seconds. 

We included three measures from the theory of \ac{CNA} that are applicable to all generic network topologies. These are as follows,
\begin{itemize}\label{item:RemovalStrat}
\item Betweenness Centrality (undirected, unweighted)
\item Degree Centrality (undirected, unweighted)
\item Closeness Centrality (undirected, unweighted)
\end{itemize}
And one measure specifically designed for power grids called \emph{Node Significance} (directed, weighted) explained previously in Sect. \ref{sec:PastWork}. 

These measures provide us with information about the importance of a node in a network with respect to the network itself. We take only undirected and unweighted \emph{traditional} measures to emphasize on the context-independent part of our study.

On the basis of the above discussion we form the following five strategies for node removal which trigger a cascading failure,
\begin{enumerate}
\item Betweenness based removal
\item Degree based removal
\item Closeness based removal
\item Node Significance based removal
\item Random removal
\end{enumerate}
1-4 belong to the targeted removals and are based on the importance of a node within a network. A random removal signifies failure or shut-down of a random component.

\subsection{Flow Redistribution}
Once a node has been removed, flows start to redistribute in the network. When a node is removed from the network, the incoming and outgoing links have certain loads that redistribute to the neighbours of the removed node. 

\begin{figure}
\begin{center}
	\includegraphics[width = 0.8\textwidth]{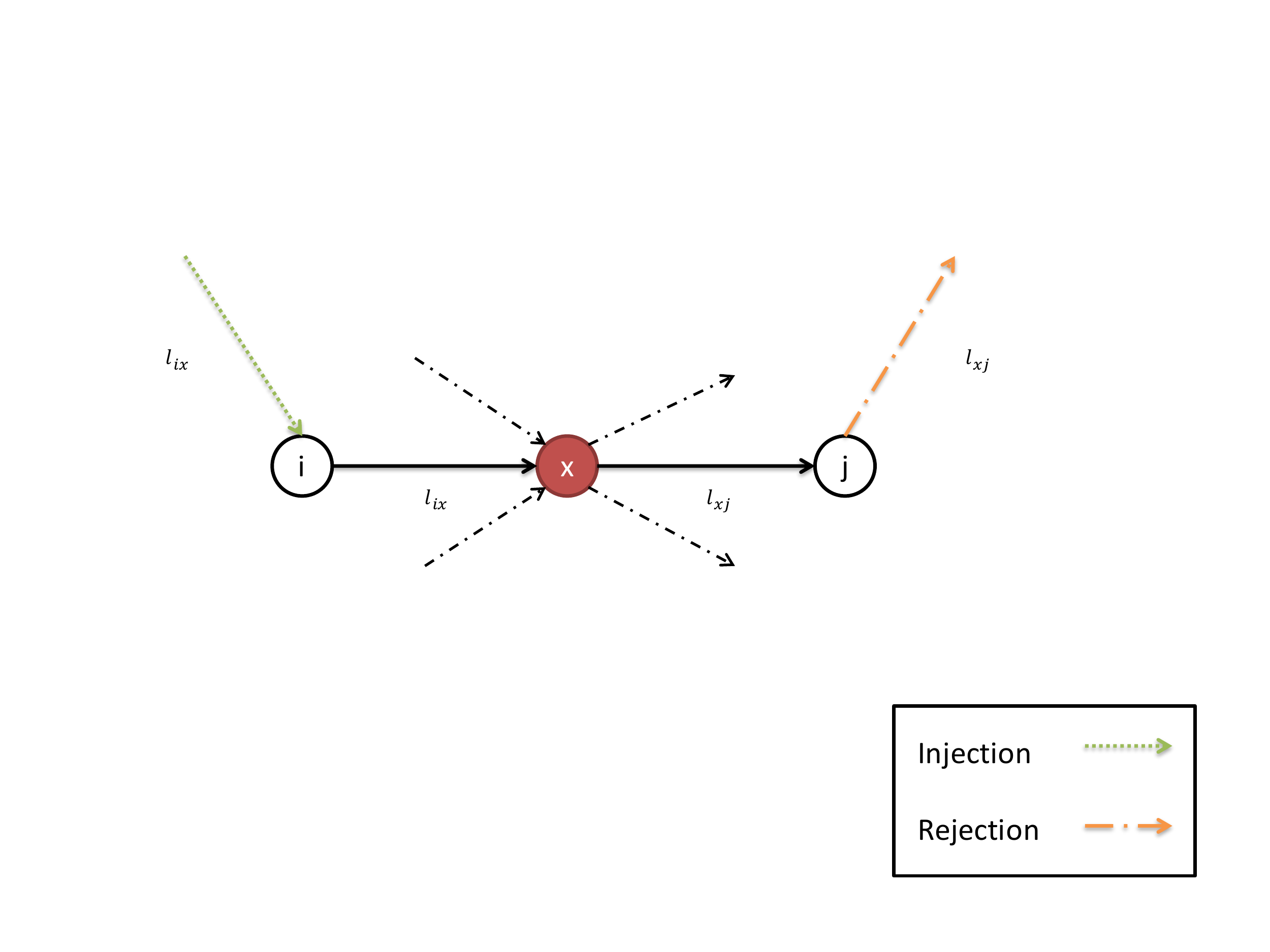}
\caption{Node N is removed from the network and flows redistribute to its neighbours}
\label{Strategy}
\end{center}
\end{figure} 

Figure \ref{Strategy} illustrates the flow redistribution mechanism that we designed for re-routing power in case of a failure or an attack on a power grid. When node $x$ is removed, all the incoming loads at node $x$ are injected to its neighbouring nodes at the other end of the transmission links. Similarly, all the outgoing loads from node $x$ are rejected from its neighbouring nodes at the other end of the transmission links. It can be seen from Fig. \ref{Strategy} that the link carrying load $l_{ix}$ will be removed from the network after node $x$ has been removed and the load $l_{ix}$ will be injected into node $i$. This procedure is applied to all the incoming neighbours $i$ of node $x$. Similarly, for all the outgoing neighbours $j$ of node $x$, a rejection of load $l_{xj}$ occurs at each node $j$. 

An injection makes sure that the load that was going into node $x$ will now be redistributed to other parts of the network and a rejection allows for removing the excessive load from a part of the network that does not actually receive it any more. 

A cumulative shift in load at each node is graphically shown in the flow redistribution part of the model in Fig. \ref{fig:Model}. Due to the numerous injections and rejections, loads at each node either increase or decrease and some might even change the direction of flow of power.  

Using \eqref{eq:PTDFMatrix} we can deduce equations formulating injections and rejections for removal of node $x$ as follows,

\begin{equation}\label{eq:Injection}
I = \sum_{i \in IN} l_{ix} \cdot u_i \quad ,
\end{equation}
\begin{equation}\label{eq:Rejection}
R = \sum_{j \in ON} l_{xj} \cdot u_j \quad ,
\end{equation}
where $IN$ and $ON$ are sets of incoming and outgoing neighbours of $x$ and $u_i$ and $u_j$ are $K \times 1$ column vectors specifying all row elements of the $i^{th}$ and $j^{th}$ column of the PTDF matrix, respectively. $I$ and $R$ are $K \times 1$ column vectors of cumulative injections and rejections at every transmission link of the power grid due to removal of node $x$. 

Using \eqref{eq:Injection} and \eqref{eq:Rejection}, we can state the following,

\begin{equation}\label{eq:BranchLoads}
L_{NEW} = L_{OLD} + I - R \quad ,
\end{equation}
where $L_{NEW}$ and $L_{OLD}$ are $K \times 1$ column vectors specifying the loads at each transmission link. 

After the initial trigger for simulating cascading failures, wherein we remove a node and all the incoming and outgoing links of \emph{this} node from the network, the flow dynamics of the power grid change. We mark these nodes and links as out-of-service. As the next step, the model calculates the new loads for all the in-service transmission links. The loads flowing through each of the remaining nodes of the network are calculated by simply summing up all the incoming or outgoing loads of each node. 

Each node may or may not have exceeded its capacity due to the possibility of extra incoming load. In \eqref{eq:LoadCapacity} if the boolean statement $S$ is true then at time $t$ node $i$ has exceeded its capacity and will be made out-of-service together with all its incoming and outgoing links. 

\begin{equation}\label{eq:LoadCapacity}
S = 
\begin{cases}
	1& \text{ if } L_i(t)>C_i \, \\
	0& \text{ otherwise } \,
\end{cases} ,
\end{equation} 

\begin{figure}
\begin{center}
	\includegraphics[width = .8\textwidth]{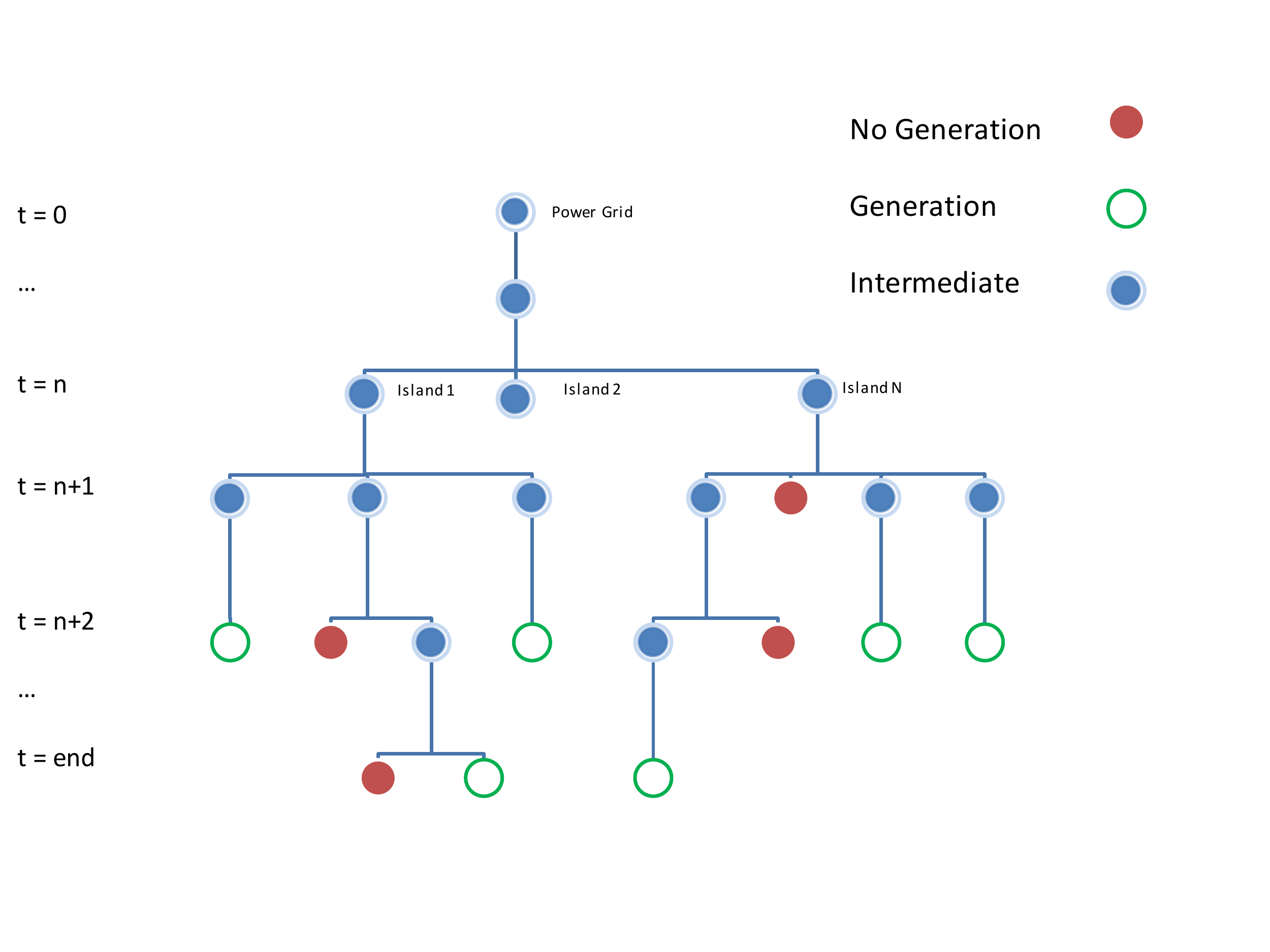}
\caption{A tree structure depicting the logical depth of the model and iterating procedure}
\label{fig:Tree}
\end{center}
\end{figure}

As a result of rendering some nodes and their links out-of-service, some parts of the network may get disconnected and disintegrate into separate islands. Each island of the disintegrated network may or may not have generation substations. If an island is free of any generating source then it is dead, marked red in Fig. \ref{fig:Tree}. However, if an island has a generating source, then we iterate the model on all such islands by removing all the nodes (together with their connected links) with the value of $S$ as $1$ or \emph{true}, ie. overloaded nodes. Each island may result in an additional group of islands and the model is iterated on each group of islands belonging to the same time step simultaneously. This procedure can be encapsulated by a tree structure illustrated in Fig. \ref{fig:Tree}. 

Some islands may not be affected completely by flow redistribution, i.e. they still have a power generating source. In this case they are still functioning and are marked green in Fig. \ref{fig:Tree}. Cascading effects have subsided in such islands and they are still providing energy for a part of their consumers. All the intermediate islands that could break down completely due to cascading failure are marked blue. The tree structure shows how an island (original power grid under consideration) can undergo changes in the flow of power and disintegrate into several islands some of which are functioning and some are deeply affected by the cascade.  

At each time step of the simulation, a few nodes may get overloaded. The transition between simulation steps is assumed to be static as compared to a dynamic transient analysis where power flows mimic the real time taken for power to flow through a line with certain physical properties. Overloaded nodes are kicked out of the simulation all at once for any intermediate state of the simulation. 

\subsection{Output}
We carry out simulations to observe the evolution of a cascade. The output of our model consists of several islands that may or may not be able to deliver power to their customers. At each time step a network is produced consisting of one or more disconnected components (islands). Every island contains information about the increase or decrease in load of each node belonging to that island. This group of disconnected islands is fed back as input to the model for the next time step after removing the overloaded nodes and connected links from the network. The flow redistribution model is iterated again on these group of islands. After processing this information from each time step we are able to assess the damage caused to the power grid. 

\section{Results}\label{sec:Results}
We followed a simulation based approach to solve our research problem. Simulations require an environment and the settings for that environment are defined in the next section followed by the approach used in the simulations and the discussion that stimulated from the results of our simulations.

\subsection{Scenario}\label{sec:Scenario}
The scenario for carrying out our simulations is composed of three parts, 
\begin{itemize}
\item \underline{Network Model}\\
The main focus is on the high-voltage European power grid. This data consists of the topology and load settings of the grid and is available on-line from the database of \ac{UCTE} \cite{UCTE}. Apart from this we have also considered IEEE power system models (N = 30, 39, and 118)for comparison of a real power grid to synthetic network topologies.
\item \underline{Node Removal Strategies}\\
These strategies are based on centrality measures for nodes in a network. We focus on betweenness centrality, closeness centrality and degree centrality \cite{Newman2} as theoretical measures from \ac{CNA} to compare with node significance \cite{Koc}. We also analyse the network using an average of a hundred random removals. The distinction will be made wherever necessary. 
\item \underline{Tolerance Parameter}\\
We have taken a realistic range of the tolerance parameter, $\alpha \in [1.01,2.8]$ \eqref{eq:Alpha}, in various settings. In reality this is different for each link/node of a network model that represents a power grid but for simplicity we have assumed a homogeneous loading level of the network, i.e. each node is loaded to the same percentage of its capacity.
\end{itemize}

\subsection{Performance Measures}\label{sec:Perf}
The following performance measures are used to assess the vulnerability of power grids,
\begin{itemize}
\item \underline{Damage 1 ($D_1$)}\\
The fraction of nodes being overloaded at any time during a cascading failure represents the cumulative damage caused to a power grid at that point in time. This metric as a performance measure is very relevant to our work because a cascading failure propagates as a result of \emph{overloading}. 
\item \underline{Damage 2 ($D_2$)}\\
The fraction of energy demand that cannot be matched at any time during a cascading failure represents the cumulative damage caused to power grid users (consumers) at that point in time. This metric is relevant to our work because cascading failures directly affect the \emph{consumers} of a power grid. 
\end{itemize}

\subsection{Simulations}
We simulated our cascading failure model defined in Sect. \ref{sec:Model} using proposed flow redistribution mechanism on the \ac{UCTE} network model and other IEEE synthetic topologies. Cascading effects were observed for several removal strategies under different input parameter settings. To encapsulate a wide domain of removal strategies we removed nodes with the highest betweenness centrality, degree centrality, closeness centrality and node significance separately and let the network mimic redistribution of flows that lead to a power outage if at all. For random failures and faults in the network, we removed a random node and averaged the damage caused by a hundred such removals. We used previously defined measures of \emph{damage} to quantify the vulnerability of a power grid to cascading failures. We varied the tolerance parameter $\alpha$ and observed the evolution of $D_1$ and $D_2$ caused to the network. Note that each iteration of the simulation is counted as 1 time step. 

\begin{figure}
\begin{center}	
	\subfloat[Fraction of overloaded nodes]{
	\includegraphics[width=.8\textwidth]{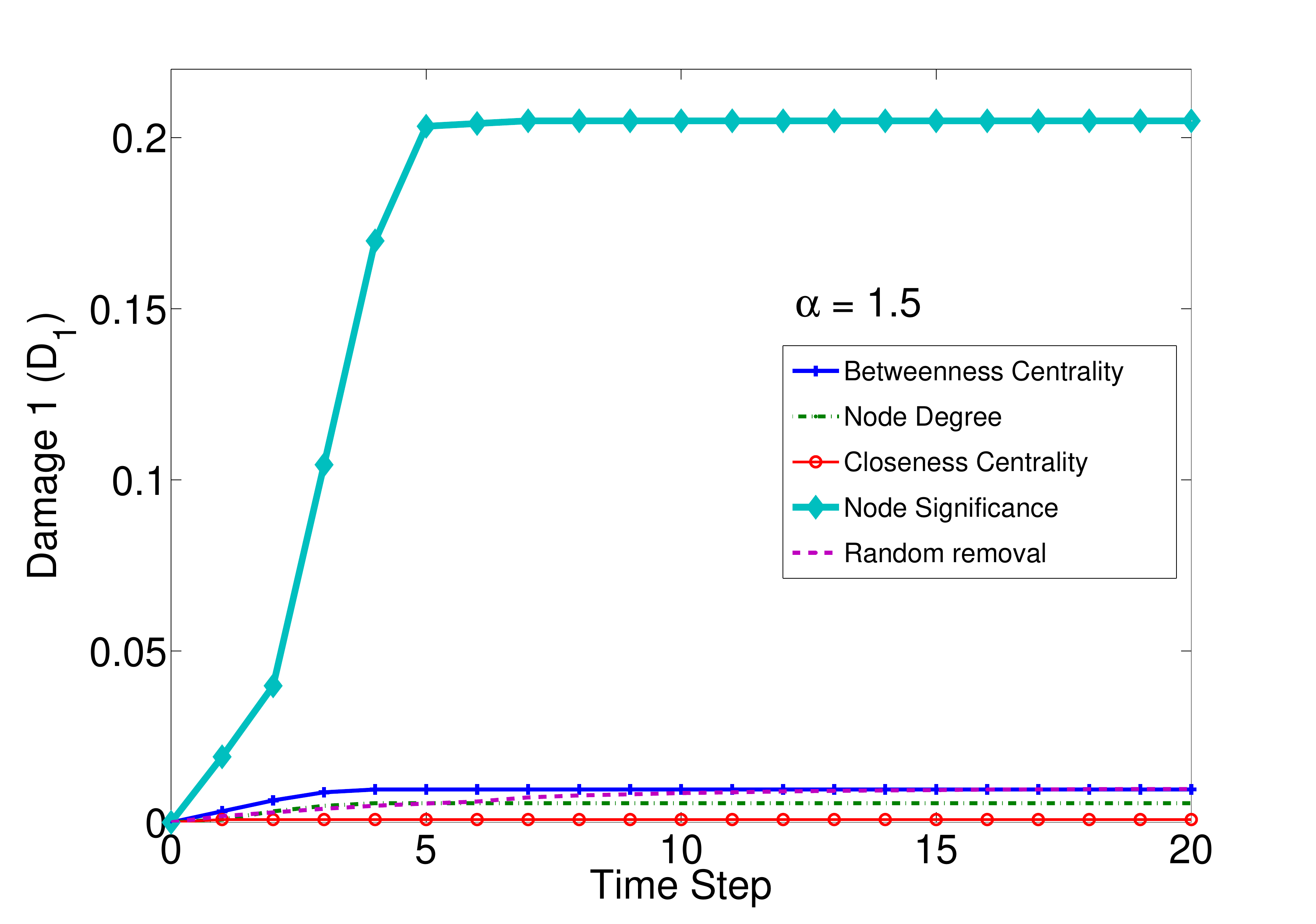}
	\label{fig:figure1}
	}
	
	\subfloat[Dissatisfied energy demand]{
	\includegraphics[width=.8\textwidth]{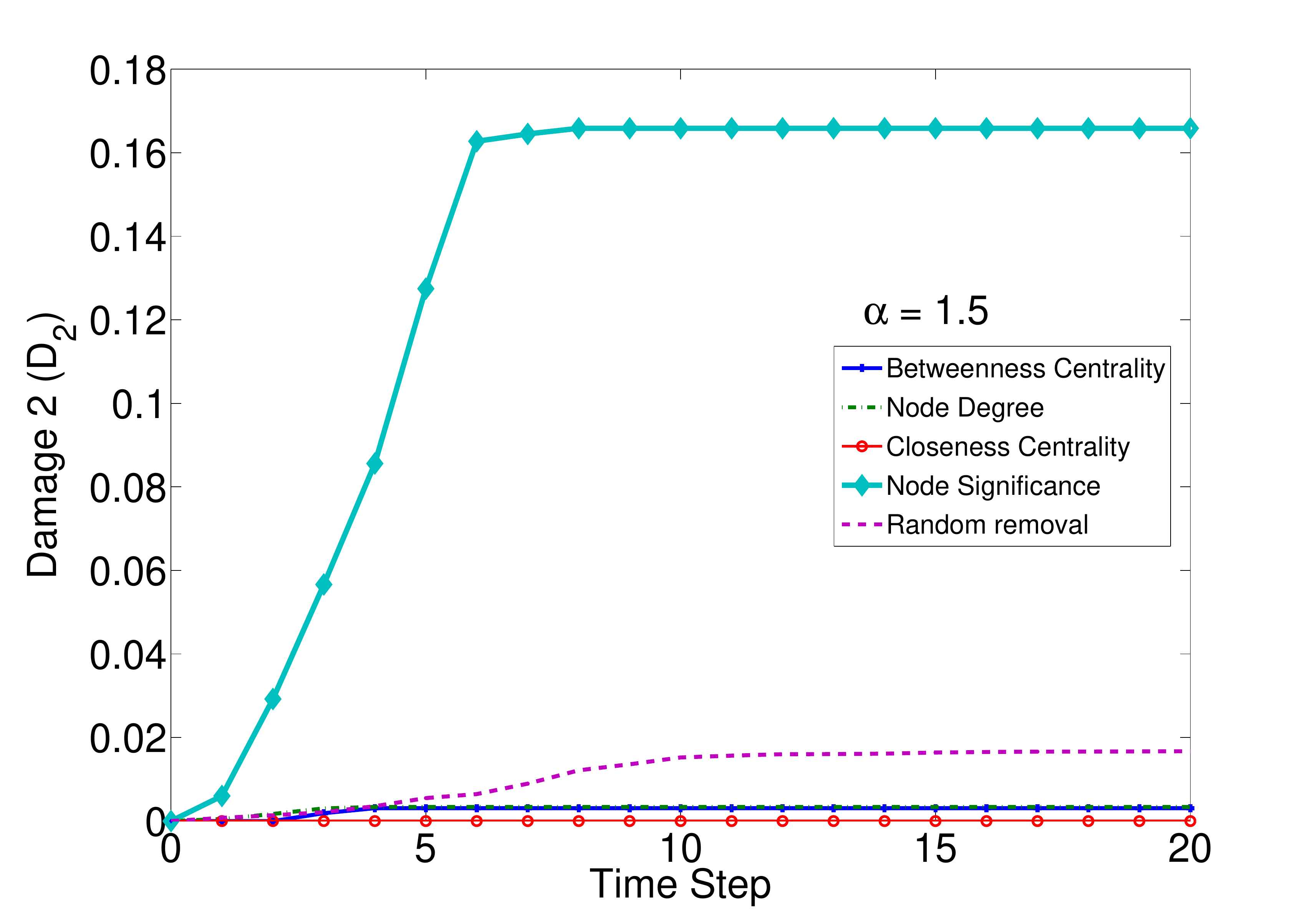}
	\label{fig:figure2}
	}
\caption{Comparison of damage caused over time for different removal strategies due to a cascading failure for $\alpha = 1.50$}
\label{fig:Alpha1p5}
\end{center}
\end{figure}

\subsection{Discussion}
Figure \ref{fig:Alpha1p5} shows a comparison of the damage caused by the five different removal strategies (of Subsect. \ref{sec:Scenario}), using the two ways to define damage (Subsect. \ref{sec:Perf}) for a realistic loading level ($\alpha = 1.5$). We observe that the damage increased with time and after a certain point in time the effects stabilised. The two performance measures quantifying the damage (one in Fig. \ref{fig:figure1} and the other in Fig. \ref{fig:figure2}), show a similar trend. Therefore our conclusions do not depend on the chosen performance measure. As illustrated in Fig. \ref{fig:Alpha1p5}, the damage caused by a removal strategy based on node significance results in approximately $20$ times more damage compared to removal strategies based on the theory of \ac{CNA} (Sect. \ref{item:RemovalStrat}). The attack strategies based on the traditional centrality measures even show less damage than random node removal does.

This shows that measures from \ac{CNA} underestimate the vulnerability of a power grid to targeted attacks. Apparently, the removal strategies based on these measures do not attack the most important nodes in the grid. Therefore, these measures are not the best way to comment on the centrality of nodes in a power grid. Heuristically speaking, \emph{node significance} may be an upper bound for the worst case damage caused to a power grid due to cascading failures. In order to focus on this worst case, i.e. the most disruptive removal strategy, we further concentrate on node significance.
 
\begin{figure}
\begin{center}
\includegraphics[width=.8\textwidth]{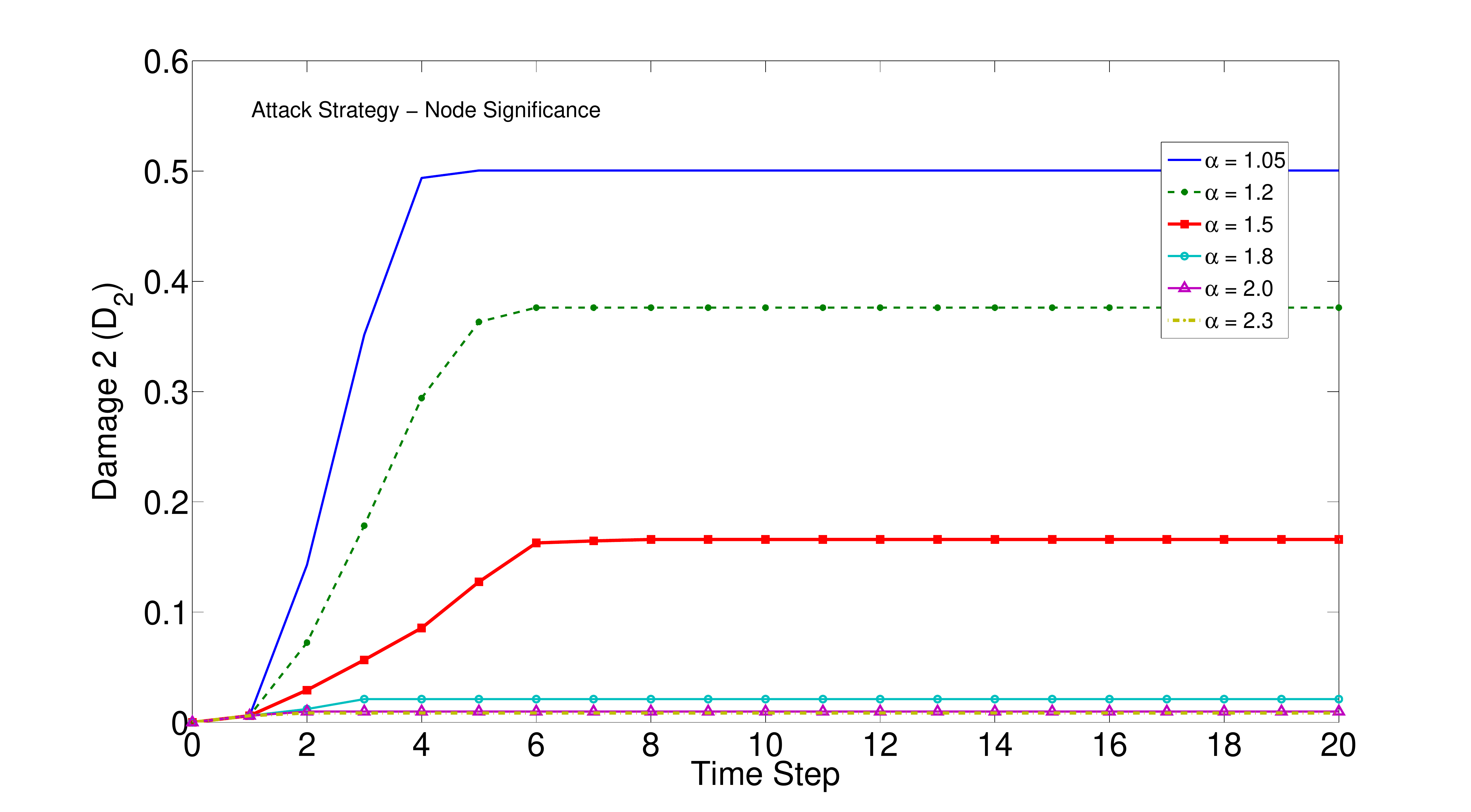}
\caption{Effect of load tolerance $\alpha$ on the damage caused to the system under a targeted attack based on node significance}
\label{fig:EnergyDamageNS}
\end{center}
\end{figure}

\begin{figure}
\begin{center}
	\subfloat[UCTE Network]{
	\includegraphics[width=.6\textwidth]{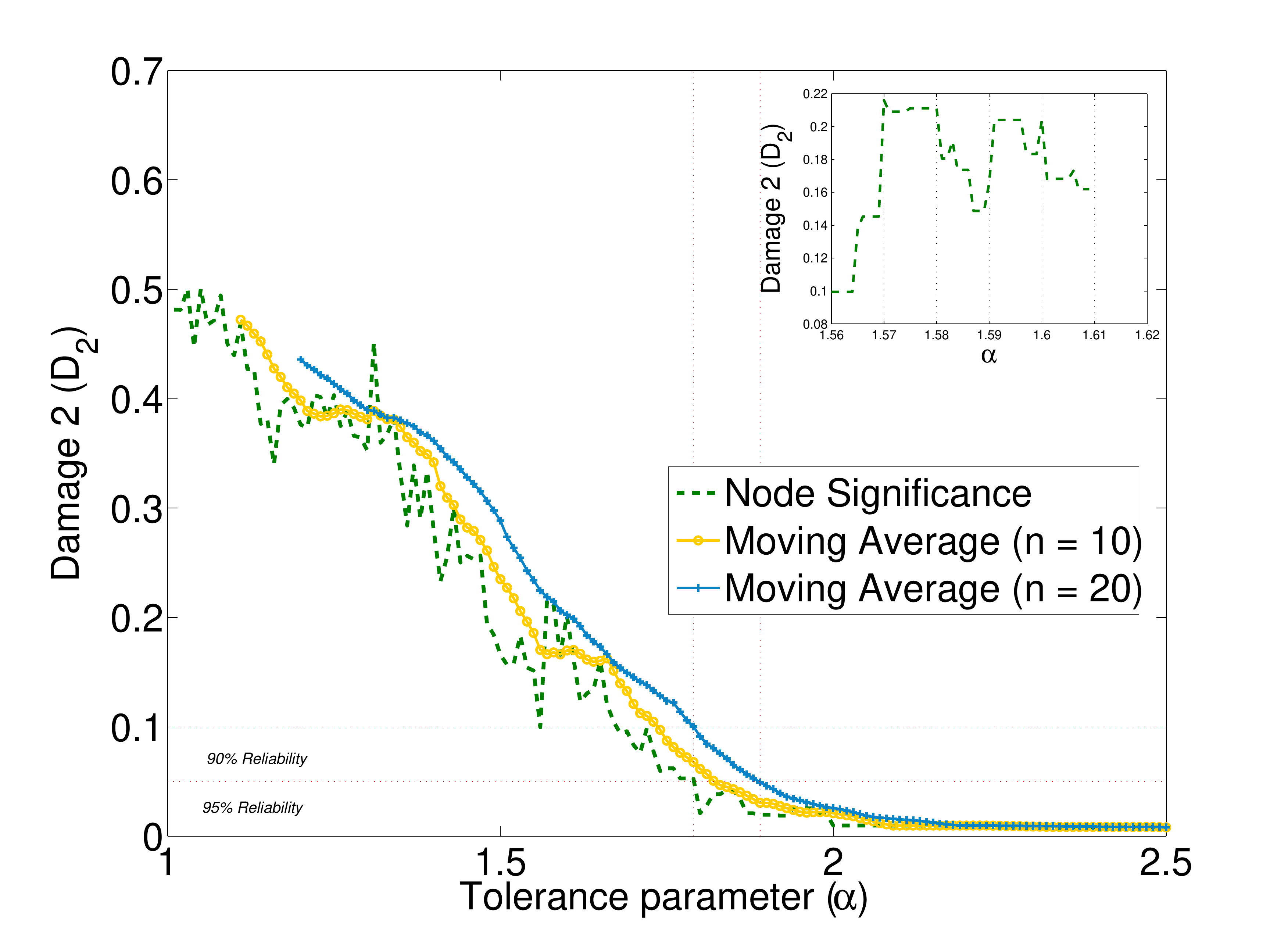}
	\label{fig:EnergyAlphaNS1}
	}
	\subfloat[IEEE Synthetic topologies]{
	\includegraphics[width=.6\textwidth]{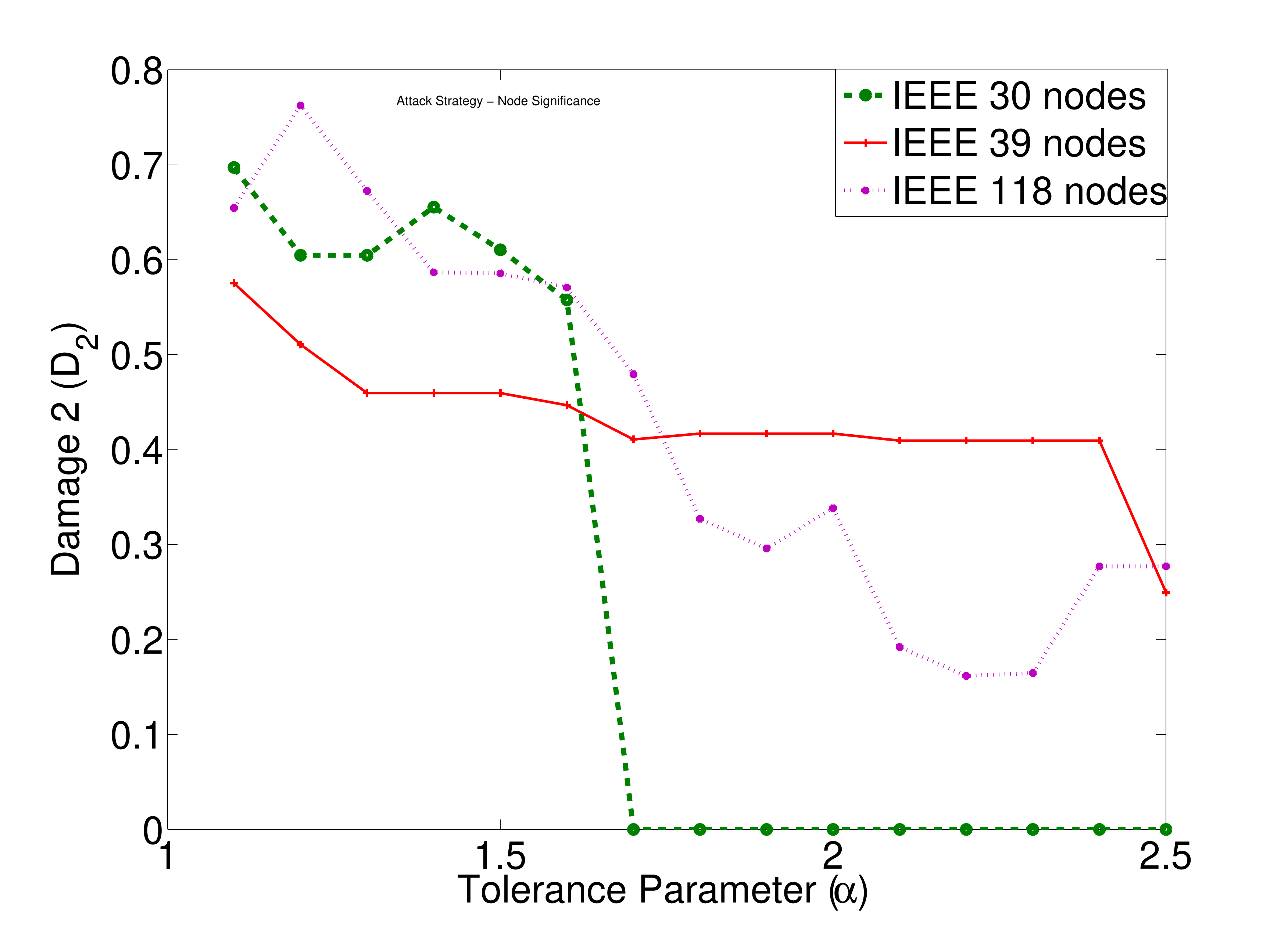}
	\label{fig:EnergyAlphaNS2}
	}
\caption{Final $D_2$ caused by removing the most significant node in the network for different values of tolerance parameter. The dotted lines in (a) represent reliability level of the system after a cascading failure has subsided, i.e. $90\%$ reliability ($10\%$ damage) $95\%$ Reliability ($5\%$ Damage). }
\label{fig:EnergyAlphaNS}
\end{center}
\end{figure}

Since both $D_1$ and $D_2$ plotted over time for a particular value of $\alpha$ are comparable to each other, we use only $D_2$ for expressing further results, i.e. fraction of unavailable energy. As expected, in Fig. \ref{fig:EnergyDamageNS}, we see that the final damage after a cascading failure has subsided decreases with an increase in the tolerance parameter. The question now comes up how large the tolerance parameter should be in order to limit the damage to a given value. We address this question in Fig. \ref{fig:EnergyAlphaNS} by plotting the total \emph{damage} after cascading failure has subsided as a function of the tolerance parameter. It can be seen that the damage caused by cascading failures decreases by increasing the tolerance parameter, however, it is not a monotonically decreasing function. It behaves erratically for a small change in the tolerance parameter. 

The reason for this erratic behaviour lies in the origin of flow redistribution. Each sub-station of the power grid is assumed to be homogeneously loaded at the beginning. Loads at each substation change after a failure or shut-down causes flows to redistribute. As a result each sub-station has a different threshold for breaking down. To illustrate this we take $\alpha = 1.56$ and remove a node with the highest node significance from the network. There are $21$ nodes that get overloaded in the first time step. At $\alpha = 1.57$ there are 20 nodes that get overloaded in the first time step. Since the flows start redistributing after the first time step, this change in the number of overloaded nodes restructures the generation of islands and the redistribution takes different paths in different islands for these two values of $\alpha$. Since we do not have reversible nodes, islands are generated because connectedness of the network cannot be kept intact after overloaded nodes go out-of-service. Therefore, the dynamics of flows change considerably for a very small change in the value of $\alpha$ (step increments of $.01$) as shown in Fig. \ref{fig:EnergyAlphaNS1}. We change the step increments to $.001$ and the inset of Fig. \ref{fig:EnergyAlphaNS1} shows that the damage curve becomes more erratic. To compare our results for a more generic overview, we plotted the final damage as a function of the tolerance parameter also for IEEE synthetic topologies. Figure \ref{fig:EnergyAlphaNS2} confirms our approach as in showing similar patterns of degradation in damage when capacity increases ($\alpha$ increases) for different topologies of synthetic IEEE networks.

To show the long-term trend we plotted a \emph{simple moving average} of the damage caused at each step of the tolerance parameter. Simple moving average is the mean of previous $z$ discrete data points. The curve represents the general long-term trend of reducing damage to a power grid by increasing the tolerance parameter. In Fig. \ref{fig:EnergyAlphaNS1} we have taken $z = 10$ data points and $z = 20$ data points. If we have more data points in the window of moving averages, we get a smoother relation between $D_2$ and $\alpha$ owing to the definition of moving averages itself. Hence, moving average is a good numerical approximation to show the intuitive relation between the destruction caused by a power outage and maximum capacity of power system components. 

As can be seen in Fig. \ref{fig:EnergyAlphaNS1}, if we want to limit the damage to the European high-voltage power grid to $5-10 \%$ ($90-95 \%$ reliability), a value of $\alpha = 1.79 - 1.89$ for the tolerance parameter would suffice. This implies that if each sub-station of the \ac{UCTE} power grid is loaded to at most $55.8 \%$ \eqref{eq:LoadingLevel} of its total capacity, the power grid will provide energy for at least $90 \%$ of its consumers, even if an attack targeted to the most vulnerable node is carried out.

\section{Conclusions}\label{sec:Conclusions}

\subsection{Summary}

We used a simulation based approach to implement a flow redistribution mechanism that takes into account the underlying flow characteristics of power grids. We continued the research by analysing the vulnerability of the European high-voltage power grid using our enhanced model. This analysis involved a study of the damage to the grid caused by a node removal based on the highest centrality for four centrality measures: betweenness, closeness, node degree and node significance. We carried out a comparison based on these four removal strategies and an average of a hundred random removals. In the analysis `damage' is measured in terms of number of overloaded nodes and the percentage of energy demand that could not be satisfied. 

As expected, the removal of a random node caused less damage than a targeted attack for node significance. Surprisingly, this was not the case for the other centrality measures. Our studies show that the theoretical centrality measures (betweenness, closeness and node degree) underestimate the vulnerability of power grids to cascading failures in comparison with node significance, which is a context based measure. Removing a node with the highest node significance causes remarkably more damage than removing a node with the highest centrality (betweenness/closeness/degree). This shows that node significance is more suitable than the traditional centrality measures for finding the most vulnerable nodes in a power grid. 

A second result that we obtained by using our model for flow redistribution is the quantification of the needed tolerance level in the European high-voltage power grid in order to limit the damage caused by targeted attacks. If approximately eighty percent of spare capacity is available for each node (with respect to the typical node loading level), then a removal of the most vulnerable node (that is the node with highest node significance) causes at most ten percent of damage (measured in amount of dissatisfied energy demand).

By modelling flow redistribution in a realistic manner, we have developed a way to identify the vulnerable nodes in a power grid and to determine the node capacity that is needed to guarantee a certain level of robustness against targeted attacks. In addition, we showed that centrality of nodes in power grids should not be defined based on topological network properties only; the power flowing through the grid should be considered when defining centrality in the context of power grids.

\subsection{Future Work}
We examined cascading failures in power grids and wrote a detailed comparison between different centrality measures. A logical next step is to expand the typography and include more centrality measures related to link capacities and optimised flow distribution. A detailed investigation in entropy based centrality measures will also be helpful in understanding the vulnerability of power grids. 

Furthermore, transient dynamic analysis of power flow can be explored for a more real-time behaviour. So far we have assumed that transition from one simulation stage to the next is discrete. Using the temporal behaviour of power flow through transmission lines will be a step closer to reality. 

We also assume that the links are loaded homogeneously. An effort towards using real capacities of power components instead of estimates with tolerance parameter would provide for a more accurate study.

Concerning applications to the industry, with minor changes to input parameters and a suitable replacement for power distribution this model can be applied to low voltage segments to investigate their structural vulnerabilities as an initiative towards \emph{smart-grids}. 

\subsection*{Acknowledgements}
This work is funded by the NWO project RobuSmart: Increasing the Robustness of Smart Grids through distributed energy generation (a complex network approach), grant number 647.000.001.

\bibliographystyle{agsm}
\bibliography{Bibliography}

\end{document}